# Space-time areas and hadronization studies

S. Abachi*, C. Buchanan

*Department of Physics and Astronomy, University of California, Los Angeles, CA 90095, USA*

---

**Abstract**

Space-time area laws play a fundamental role in modeling strongly interacting phenomena, such as in relativistic string models, QCD on the lattice, and in hadronization modeling. For the latter, an exact expression for relevant space-time areas related to the final state hadronic events for hadrons of any mass is calculated. Usage of this expression in the UCLA hadronization model helps validate its space-time area law ("STAL") principle and unifies the treatment of heavy and light hadrons via a single fragmentation function.

*PACS:* 13.87.Fh

---

*Keywords:* Hadronization, Fragmentation, Non-perturbative QCD, Area law

* Currently at Physics Department, CSULA

* *E-mail address:* sabachi@calstatela.edu

## 1. Introduction

Space-time area laws are suggested in several fields of strongly interacting phenomena. In one example, according to the string model, hadrons behave like strings with quark quantum numbers attached to their ends [1]. This simple assumption has led to interesting results for hadronic phenomena. The action of a relativistic string is proportional to the space-time area, $A$, spanned by the string; i.e., $S = \kappa A$, where $\kappa$ is the string tension. However, as in Fig. 1(a), energetic strings stretch and may break repeatedly by creation of quark-antiquark pairs at new endpoints, producing new string segments. This may continue until stable hadrons appear, as shown in Fig. 1(b). For example, with the assumption that P is a universal probability per unit length and time that a string may break anywhere along its length, Artru and Bowler [2] show that the probability for a particular string configuration is $exp(-PA)$. And when this is combined with other simple assumptions, it leads to the event probability of $exp(-PA_o)$, where $A_o$ is the space-time area swept out by a particular classical event configuration.

It is also interesting to note that in one-space and one-time dimensions, it has been shown that the string model is equivalent to chromodynamics [3].

QCD on the lattice results imply an area law as well. Such properties can be seen in the well known Wilson loop integral expressions [4] dominating the non-perturbative QCD interaction results.

Perturbative QCD, applied to quarks and gluons, is a valid approach at short distances. In the high coupling regime at large separations, this treatment fails as QCD becomes strongly interacting, leading to confinement and formation of colorless hadrons from colored partons. Hadronization is therefore a non-perturbative process that is not yet developed nor understood from fundamental theoretical principles. Therefore, for studying the hadronic events produced in high energy collisions, only several types of phenomenological models are currently available, most of which are successfully developed and implemented in the form of Monte Carlo programs. Traditionally, the underlying bases for these models are not QCD motivated and consist of several different approaches, such as the independent fragmentation modeling of Field and Feynman [5], Hoyer [6], and Ali [7]; the string fragmentation approach by Lund [8]; cluster fragmentation methods of Webber [9]; and other approaches with some variations [10].

Although at this time there is no analytical scheme that can begin with a QCD Lagrangian and predict the existing hadronization's experimental picture, it is widely assumed that the relevant underlying physical process for hadronization essentially involves non-perturbative QCD. Therefore it is important to make as much connection as possible with this regime in modeling of this phenomenon. Currently, the general state of most models may be summarized as follows:

(i) They all do a remarkable job in the perturbative part (where Parton showers dominate) using standard QCD perturbation methods. This is the part that is essentially responsible for the event topology, described by parameters such as sphericity, aplanarity, thrust, etc. While accurate results are predicted for the final state hadronic rates and spectra, the non-perturbative regime of the hadronization process is phenomenologically modeled, and the modeling of this important part effectively begins with a parametrized fragmentation function.

(ii) Most models use one fragmentation function to account for the light hadrons containing light primary quarks, while other functional forms are used to describe the heavy quarks case. There is often a discontinuity in going from light to heavy hadrons.

Some time ago, the UCLA studies considered a somewhat different approach which may address some of the forgoing issues [11]. The outcome of these studies can be viewed as a first step toward making a reasonable connection with hadronization's true non-perturbative QCD basis while unifying the treatment of light and heavy hadrons in its implementation. In a short review, it will be shown that an important space-time area emerges from this approach for which an exact derivation will be given, followed by a summary of comparisons with data after this area expression is utilized.

## 2. Nature of hadronization

As a common beginning of a hadronization process, consider the example of a high energy annihilation from an $e^+e^-$ collision. As depicted in Fig. 1(a), a $\gamma^*/Z^o$ is produced from the annihilation which decays to a primary quark-antiquark pair, $q_o\bar{q}_o$ − completely described by the electro-weak theory. This primary pair quickly fragment and dress themselves as jets of hadrons; thus the word "hadronization" (Fig. 1(b)). Such processes can normally be formulated and/or conceptually realized more conveniently in the context of relativistic strings. The quark and antiquark in the pair fly apart, stretching a narrow color flux tube between them. Initially, when distances are less than a fraction of one Fermi, perturbative interactions dominate and a Parton shower develops which can be calculated by standard QCD perturbation methods. At larger distances perturbative QCD fails, and it becomes energetically more favorable for the string to fragment by breaking repeatedly and ultimately end up with several stable colorless yo-yo mode strings called hadrons. If at the break points quark-antiquark pairs are created, mesons will be produced, while diquark-antidiquark pair creation can lead to baryon formation. However, although there are some differences in the baryon implementation [11], the formulation is identical for both types. Therefore for simplicity, only the example of mesons is used for the derivations in this manuscript.

Although the last non-perturbative steps leading to final state hadrons are successfully modeled by various types of fragmentation schemes, the emphasis here will be on the UCLA approach which offers a QCD-motivated basis.

### *2.1. UCLA scheme – light hadrons*

The UCLA approach, whose details are reported elsewhere [11], has successfully identified one possible QCD-motivated path to hadronization data which may be summarized as follows: Strongly interacting QCD processes such as those described by Wilson loops and QCD results on the lattice lead to a space-time area law ("STAL") in Euclidian space. It is presumed that this law continues to STAL in Minkowski space. (See [11] for further details.)

Since the hadronization phenomenon is virtually a non-perturbative QCD process, it is natural to try taking the STAL as the single dominant physical basis for it. For example, in a high energy collision leading to a primary quark-antiquark pair, STAL simply postulates that the probability of occurrence of a particular hadronic event from such a primary pair is proportional to the negative exponential of the area in space-time swept out by the event – that is, $exp(-b'A_{plane})$, where $A_{plane}$ is the entire 1 + 1 dimensional area occupied by the event, an example of which is shown in Fig. 1(b). When this assumption is combined with the conservation of energy-momentum, an event weight function is obtained. Next, from this event weight function a fragmentation function is derived. The following UCLA fragmentation function ("UCFF") applies to light hadrons containing light primary quarks [11]:

$$f(z)=\frac{NC^2}{(4\pi)^2}\frac{(1-z)^a}{z}\left(1-\frac{m_h{}^2}{Sz}\right)^a e^{-b'a_{xt}} \qquad (1)$$

which describes the probability density for producing a light hadron with mass $m_h$ taking a fraction $z$ of the light-cone momentum ($p^+=E+p$). In this expression, $a$ & $b'$ are arbitrary parameters arising naturally from the UCLA approach; $C$ is the Clebsch-Gordan coefficient to combine the flavor and spin of the quark and antiquark into the hadron; $N$ is a spatial "Knitting Factor", experimentally ~(2.7 fm)$^2$, to combine the quark and antiquark into the hadron's spatial wave-function; $S$ is $E_{\text{c.m}}^2$; and the light-cone momentum fraction $z$ is defined by

$$z \equiv \frac{p_h^+}{p_q^+} \tag{2}$$

where $P_h^+$ and $p_q^+$ are the light-cone momenta of hadron and primary quark respectively. For the light quarks, $a_{xt}$, appearing in the exponent of (1) is a rectangular event-related space-time area formed by the segments of the light primary quark $q_o$ world-line and the $\overline{q}_1$ world-line and its extension, an example of which is shown in Fig. 1(b) in 1 + 1 dimensions. All light quark paths are on the light-cone, and as will be shown by (24) below, $a_{xt} \sim m_h^2/z$, which introduces a natural mass suppression for heavy hadrons through the exponent of the fragmentation function. In terms of $m_h^2/z$, the fragmentation function (1) is

$$f(z) = \frac{NC^2}{(4\pi)^2} \frac{(1-z)^a}{z} \left(1 - \frac{m_h^{\ 2}}{Sz}\right)^a e^{-b\frac{m_h^{\ 2}}{z}} \tag{3}$$

This fragmentation function form is almost identical to the Lund Symmetric Fragmentation Function ("LSFF") [8], with the exceptions of the small correction term, $m_h^2/Sz$, and the important absolute normalization for any flavor-spin combinations given by $NC^2/(4\pi)^2$. It is interesting that for the case of light hadrons, the LSFF (derived based on the Lund requirement for a symmetry in implementation [8]) is almost identical in form to the UCFF (derived based on the UCLA STAL assumption).

It should be clarified that the Lund group have been aware that their fragmentation function – derived based on a symmetry requirement – carries a space-time area dependent exponent that is suggestive of an area law [8]. Beginning with other assumptions, others such as Bowler [12] and Artru-Mennessier [13] have also seen such dependences. Although motivated by these observations, the UCLA approach differs, in the sense that it begins with an area law as a single fundamental assumption, and derives the hadronization fragmentation function and the rest of the model structure [11].

Furthermore, in their "outside-in, one hadron at a time" implementation approach, both the Lund and the UCLA models [8, 11] utilize a recursive relation that requires the energy of the remaining system to be large on the hadronic mass scale after any one final state hadron is created. The validity of these fragmentation functions is only approximate at lower c.m. energies, such as at those near threshold for a heavy meson production. However, in the UCLA scheme the additional factor of $(1 - m_h^2/Sz)^a$ in (3) – which arises naturally from the UCLA approach – tends to soften the effect of this approximation (see [11] for derivations leading to this factor). This is one non-negligible contributing factor for the successful agreements with heavy meson data down to very low center of mass ("c.m.") energies.

### *2.2. UCLA scheme – hadrons of any mass*

The transition to heavy hadrons containing heavy primary quarks leads to a different exponent, as the space-time area appearing in the exponent of the fragmentation function (1) no longer remains rectangular in shape. In general, for heavy primary quarks of any mass $a_{xt} \to A_{xt}$, where as shown in Fig. 2, $A_{xt}$ is the hatched area bounded by the curved heavy quark path and the $\bar{q}_1$ world-line and its extension. Therefore, the most general UCLA fragmentation function (UCFF) valid for any hadron flavor containing a primary quark of any mass is

$$f(z) = \frac{NC^2}{(4\pi)^2} \frac{(1-z)^a}{z} \left(1 - \frac{m_h^{\ 2}}{Sz}\right)^a e^{-b' A_{xt}} \tag{4}$$

Therefore, calculation of the space-time area $A_{xt}$ is essential to the UCLA approach. Of course for the light quarks, at the limit of the quark mass $\mu \to 0$, one should have $A_{xt} \to a_{xt} \sim m_h^2/z$.

## 3. Calculation of $A_{xt}$

Since the values of the space-time area $A_{xt}$ in (4) is important to proper STAL treatment, and since $A_{xt}$ is somewhat sensitive to approximations, after some simple ground work in the next section an exact calculation of $A_{xt}$ is presented.

### *3.1. Ground work from Light quark dynamics*

Consider the relativistic string model for a $q\bar{q}$ system [1,8,13] in two dimensions (x, $t$) and its principle assumption that a uniform narrow color flux tube (string) stretched between the two quarks leads to a linear potential for their interaction. This simple picture is quantified by the constant string tension $\kappa$ and the following covariant equations describing the motion of each quark in the pair:[1]

$$\frac{dE}{dx} = \pm\kappa \ , \qquad \frac{dp}{dt} = \pm\kappa \tag{5}$$

as in Fig. 1(a), at the ends of an energetic string, $q_o$ and $\overline{q}_o$ fly apart in their c.m. system as the potential energy stored in the string rises. Soon after, at the first vertex point a new quark-antiquark pair ($q' \ \overline{q}'$) is produced, splitting the original string for the first time. More string breaks may occur at other new vertices in an inside-out fashion as the time evolves vertically upward on the page. As will be shown in this section, all production vertices are causally disconnected and therefore time-order has no significance in the development. Therefore, for the implementation purposes, a Lund style [8,11] outside-in implementation is appropriate; namely one starts from the quark side ($q_o$, in Fig. 1(b)) and moves to the left one vertex at a time toward $\overline{q}_o$. In this picture $q_1\overline{q}_1$ creates the first vertex to be implemented. Carrying on with this order, as soon as $q_1\overline{q}_1$ pair is produced all the color field lines originally connecting $q_o$ and $\overline{q}_o$ are rearranged and go from $q_o$ to $\overline{q}_1$ and from $q_1$ to $\overline{q}_o$, leading to two new separate strings containing colorless $q_o\overline{q}_1$ and $\overline{q}_o q_1$ pairs. In Fig. 1(b), $q_o\overline{q}_1$ forms the first meson on the right side, while the remnant string repeats the division several times in a cascading fragmentation, and the formulation is the same for every step of the cascade only at a smaller energy scale and in a boosted reference frame. Up to a point, before the primary string is stretched beyond a fraction of a Fermi, the situation can be considered as perturbative, and any gluon radiation and subsequent parton shower is calculable via perturbative QCD matrix elements or coherent parton shower methods [14]. However, as the $q_o\overline{q}_o$ distance grows, perturbative methods fail before the string breaks. This is the part which should be modeled.

As depicted in Fig. 1(b), when the endpoint quarks of a string are not energetic enough and the system has a small invariant mass, confinement leads to formation of a meson. (Again, although not shown, a baryon instead of a meson can be formed if the break point involves a diquark-antidiquark pair, without affecting the derivations. Therefore for simplicity, only the example of mesons is used for the derivations in this work.)

[1] All symbols of momenta used in this manuscript are their signed components along an appropriate axis.

Figure 3 shows the world-lines of $q$ and $\bar{q}$ in a bound system in its c.m. frame (dotted lines) and in a boosted frame (dashed lines). The figure is identical in space-time or in momentum space, with the exception of the units and scale. If the *xt* coordinate of the quark and antiquark are measured from their crossing point *O*, then the energy and momentum of each quark can be traced by integrations of (5), namely in any Lorentz frame,

$$E_q = E_{oq} - \kappa x\,, \quad p_q = p_{oq} - \kappa t$$

$$E_{\bar{q}} = E_{o\bar{q}} + \kappa x\,, \quad p_{\bar{q}} = -p_{o\bar{q}} - \kappa t \qquad (6)$$

By setting $E_{q,\bar{q}} = p_{q,\bar{q}} = 0$ at the quark and antiquark turning points where they stop before moving backward, one may solve for the initial momenta and energies to get

$$E_{oq} = \kappa x_q\,, \quad p_{oq} = \kappa t_q\,, \quad E_{o\bar{q}} = -\kappa x_{\bar{q}}\,, \quad p_{o\bar{q}} = -\kappa t_{\bar{q}} \qquad (7)$$

with $t_q, x_q, t_{\bar{q}}, x_{\bar{q}}$ representing the times and positions of the turning points of the quark and antiquark, respectively. Now, usage of light-cone coordinates of

$$p^{\pm} = E \pm p\,, \quad x^{\pm} = t \pm x \qquad (8)$$

leads to

$$p^{\pm}_{oq} = E_{oq} \pm p_{oq} = \kappa(x_q \pm t_q) = \pm\kappa x^{\pm}_q$$

$$p^{\pm}_{o\bar{q}} = E_{o\bar{q}} \pm p_{o\bar{q}} = \kappa(t_{\bar{q}} \pm x_{\bar{q}}) = \kappa x^{\pm}_{\bar{q}} \qquad (9)$$

and thus, the enclosed area swept by this $q\bar{q}$ pair in half a period in *xt* coordinates ($S_{xt}$) compares with that in momentum space ($S_p$) via

$$S_p = p^{+}_{oq} p^{-}_{o\bar{q}} = \kappa^2 x^{+}_q x^{-}_{\bar{q}} = \kappa^2 S_{xt} \qquad (10)$$

Relation (10) is valid for all areas formed by quark traces. For instance in Fig. 4, which the $q_o$ side of Fig. 1(b) is shown in the light-cone momentum space ($p^{\pm}$), $A_h$ and $a_p$ are related to $a_{xt}$ and $A_h^{xt}$ of Fig. 1(b), respectively. One can therefore conveniently calculate areas either in space-time or momentum coordinate frames and use (10) to relate them.

Furthermore, consider a quark and an antiquark from two adjacent vertices rushing from rest toward one another to form a meson – such as $q_1$ and $\bar{q}_2$ in Fig. 1(b). Integration of (5) reveals that by the time they cross at some point $(x_o, t_o)$, the quark coming from the right side ($R$) will develop a momentum-energy ($p_R$, $E_R$) of $(\kappa(t_R - t_o), \kappa(x_R - x_o))$, and the antiquark coming from left ($L$) will have a ($p_L$, $E_L$) of $(\kappa(t_o - t_L), \kappa(x_o - x_L))$. Since this crossing is considered the meson's creation point, the total light-cone momentum of this hadron is

$$p_h^{\pm} = p_R^{\pm} + p_L^{\pm} = (E_R \pm p_R) + (E_L \pm p_L) = \kappa\left[(x_R - x_L) \pm (t_R - t_L)\right] \quad (11)$$

Since the hadron is on the mass shell, then

$$m_h^2 = E_h^2 - p_h^2 = p_h^+ p_h^- = \kappa^2\left[(x_R - x_L)^2 - (t_R - t_L)^2\right] > 0 \quad (12)$$

which means the two vertices have a space-like separation and are thus causally disconnected. This proves the claim made earlier in this section which allows any order in the implementation of the vertices.

Now, under a longitudinal boost

$$x' = \gamma(x - vt) \ , \quad t' = \gamma(t - vx)$$

$$p' = \gamma(p - vE) \ , \quad E' = \gamma(E - vp) \quad (13)$$

These lead to the following simple properties for the light-cone variables under a boost

$$x'^{\pm} = \gamma(1 \mp v)x^{\pm} \ , \quad p'^{\pm} = \gamma(1 \mp v)p^{\pm} \quad (14)$$

and therefore areas are Lorentz invariant, because

$$S'_{xt} = x'^+ x'^- = \gamma^2(1 - v^2)x^+ x^- = x^+ x^- = S_{xt} \quad (15)$$

Similarly, the light-cone momentum fraction as defined by (2) is Lorentz invariant, because (Fig. 4),

$$z' = \frac{p_h'^+}{p_q'^+} = \frac{\gamma(1 - v)p_h^+}{\gamma(1 - v)p_q^+} = z \quad (16)$$

In Fig. 4, the leading hadron box area, $A_h$, is

$$A_h = p_h^{\ +} p_h^{\ -} = (E_h + p_h)(E_h - p_h) = m_h^{\ 2} \quad (17)$$

Also, using (2) one gets

$$A_h = p_h^+ p_h^- = z p_q^+ p_h^- = z a_p = \kappa^2 z a_{xt} \quad (18)$$

where $a_p$ is the hatched area in Fig. 4 swept out by the massless $q_o$ in momentum space before the rightmost meson is formed. (The first crossing of the two quarks, $q_o$ & $\bar{q}_1$, is defined as the formation point of the meson.) If $A_h^{xt}$ (Fig. 1(b)) is the *xt* analog of $A_h$ (Fig. 4), beginning with (18) and using (10), one concludes

$$z = \frac{A_h}{\kappa^2 a_{xt}} = \frac{A_h^{xt}}{a_{xt}} \quad (19)$$

Again, in Fig. 3 consider the hadron rectangle with side $l$ in the $xt$ coordinate system viewed in the c.m. frame. When the two massless quarks come to rest at their classical turning points ($A$ & $B$), the potential energy of the system equals the hadron rest mass energy, namely

$$m_h = \kappa L_{\max} = \kappa d = \kappa\sqrt{2}l \tag{20}$$

where the maximum separation of $q_o$ and $\bar{q}_1$ is denoted by $L_{max}$, and $d$ is the diagonal of the square ($AB$ line segment). This gives

$$m_h^2 = 2\kappa^2 l^2 = 2\kappa^2 S_{xt} \tag{21}$$

Due to Lorentz invariance, this is valid in any boosted $xt$ frame. One now has

$$A_h^{xt} = \frac{m_h^2}{2\kappa^2} \tag{22}$$

where the area $A_h^{xt}$ of Fig. 1(b) takes the place of the generic area $S_{xt}$. In momentum space,

$$A_h = \frac{m_h^2}{2} \tag{23}$$

Also, from (19) and (22) one learns

$$a_{xt} = \frac{m_h^2}{2\kappa^2 z} \tag{24}$$

which establishes the claim made in section 2.1 that $a_{xt} \sim m_h^2/z$. This outcome, along with other results derived here, will be useful in the following section for calculation of the important space-time area, $A_{xt}$.

### *3.2. $A_{xt}$ area expression from dynamics of heavy quarks*

For heavy primary quarks the area of interest is $A_{xt}$ – the hatched area of Fig. 2. To calculate $A_{xt}$, one may begin by integrating the equations of motion of (5) for the primary quark $q_o$. For convenience, the integration is done in a boosted frame where $q_o$ is initially at rest, and thus, as shown in Fig. 4, massive $q_o$ and $\bar{q}_o$ both move in the negative $x$ direction soon after. Since at $t$=0, $p_{q_o}$=0, then integration of (5) yields

$$p_{q_o} = -\kappa t \ , \qquad E_{q_o} = \mu - \kappa x \tag{25}$$

which means that the total energy of $q_o$ with mass $\mu$ at time $t$ is

$$E_{q_o} = \sqrt{\mu^2 + \kappa^2 t^2} \tag{26}$$

Comparison of (25) and (26) yields

$$E_{q_o} = \sqrt{\mu^2 + \kappa^2 t^2} = \mu - \kappa x \tag{27}$$

or

$$\left( x - \frac{\mu}{\kappa} \right)^2 - t^2 = \frac{\mu^2}{\kappa^2} \tag{28}$$

This is a hyperbola describing the world-line of $q_o$ in the boosted frame.

The equation of motion of $\bar{q}_o$ can also be determined in the boosted frame by integrating (5), which gives

$$p_{\bar{q}_o} = \bar{p}_o^b + \kappa t \ , \quad E_{\bar{q}_o} = \bar{E}_o^b + \kappa x \tag{29}$$

where $\bar{p}_o^b$ and $\bar{E}_o^b$, the initial momentum and energy of $\bar{q}_o$ in the boosted frame, are

$$\bar{E}_o^b = \frac{S}{2\mu}\left( 1 - \frac{2\mu^2}{S} \right), \quad \bar{p}_o^b = -\frac{S}{2\mu}\left( \sqrt{1 - \frac{4\mu^2}{S}} \right) \tag{30}$$

It is also useful to define, for $\bar{q}_o$, the initial light-cone momenta in the boosted frame, namely

$$\bar{p}_o^{b\pm} \equiv \bar{E}_o^b \pm \bar{p}_o^b = \frac{S}{2\mu}\left( 1 - \frac{2\mu^2}{S} \mp \sqrt{1 - \frac{4\mu^2}{S}} \right) \tag{31}$$

Evidently, $\bar{p}_o^{b+}$ is small, because if expanded in powers of $\mu^2/S$, one gets

$$\bar{p}_o^{b+} \approx \frac{S}{2\mu}\left( 1 - \frac{2\mu^2}{S} - \left( 1 - \frac{2\mu^2}{S} - \frac{2\mu^4}{S^2} - \frac{4\mu^6}{S^3} \ldots \right) \right) \approx \frac{\mu^3}{S} + \frac{2\mu^5}{S^2} + \ldots \tag{32}$$

With the leading term being $\mu^3/S$, $\bar{p}_o^{b+}$ is small for not-so-heavy quarks and/or for large c.m. energies. On the contrary, $\bar{p}_o^{b-}$ is large and is of the order of $S/\mu$.

Continuing with the description of $\bar{q}_o$ motion and using (29), the energy variations with location and time may be written as

$$\bar{E}_{\bar{q}_o} = \sqrt{\mu^2 + \left( \bar{p}_o^b + \kappa t \right)^2} = \bar{E}_o^b + \kappa x \tag{33}$$

or

$$\left(x + \frac{\overline{E}_o^b}{\kappa}\right)^2 - \left(t + \frac{\overline{p}_o^b}{\kappa}\right)^2 = \frac{\mu^2}{\kappa^2} \tag{34}$$

which is a hyperbola describing the world-line of $\overline{q}_o$ in the boosted frame.

As shown in Fig. 5, once the primary color tube breaks, a new quark-antiquark pair, $q_1\overline{q}_1$, is generated at the vertex point *V*, leading to the creation of the primary meson containing $\overline{q}_1$ and $q_o$. These quarks are accelerated towards each other until they cross at point *E*, where the primary meson is first created. By then, $\kappa\Delta x$ field energy is transferred to kinetic energy of the quarks and $\kappa\Delta t$ momentum is acquired by them. For $q_o$ at point *E*, (25) may be written as

$$p_{q_o} = -\kappa t_E \ , \quad E_{q_o} = \mu - \kappa x_E \tag{35}$$

and the energy and momentum of $\overline{q}_1$ at this crossing point may be written as

$$p_{\overline{q}_1} = \kappa(t_E - t_V) \ , \quad E_{q_1} = \kappa(x_E - x_V) \tag{36}$$

Now, by adding the $q_o$ and $\overline{q}_1$ energies and momenta, the primary hadron's energy and momentum at its point of creation, *E*, can be found. Namely,

$$p_h = p_{q_o} + p_{q_1} = -\kappa t_V \ , \quad E_h = E_{q_o} + E_{q_1} = \mu - \kappa x_V \tag{37}$$

The light-cone momenta of the hadron are

$$p_h^+ = \mu - \kappa(x_V + t_V) \ , \quad p_h^- = \mu - \kappa(x_V - t_V) \tag{38}$$

Since the meson is on the mass shell, then

$$m_h^2 = [\mu - \kappa(x_V + t_V)][\mu - \kappa(x_V - t_V)] \tag{39}$$

In the boosted reference frame $q_o$ is initially at rest and thus at $t = 0$ it has no momentum and energy except for rest mass energy, i.e., $p_{q_o}^+ = \mu$. Therefore, from the definition of the light-cone momentum fraction given by (2), one has

$$z = \frac{\mu - \kappa(x_V + t_V)}{\mu} \tag{40}$$

If Eqs. (39) and (40) are simultaneously solved for the coordinates of vertex point *V*, then

$$x_V = -\frac{m_h^2 - \mu^2 z(2 - z)}{2\kappa\mu z} \ , \quad t_V = \frac{m_h^2 - \mu^2 z^2}{2\kappa\mu z} \tag{41}$$

The world-line of the massless $\overline{q}_1$ quark is on the light-cone and originates from the vertex point *V*. Hence, the equation of this line is

$$t - t_V = x - x_V \tag{42}$$

Replacing for $t_V$ and $x_V$ from (41), one gets

$$t = x + \frac{m_h^2}{\kappa\mu z} - \frac{\mu}{\kappa} \tag{43}$$

Since the crossing point $E$ is on this line (Fig. 5), the intersection of this line and the $q_o$ hyperbola of (28) gives the following coordinates for point $E$

$$t_E = \frac{m_h^2}{2\kappa\mu z} - \frac{\mu^3 z}{2\kappa m_h^2}\ , \qquad x_E = \frac{\mu}{\kappa} - \frac{m_h^2}{2\kappa\mu z} - \frac{\mu^3 z}{2\kappa m_h^2}$$

(44)

For convenience, now the $xt$ coordinate axes are rotated by 45° and are named $X^{\pm}$ as shown in Fig. 5. They are related to the $xt$ system by

$$X^{\pm} = \frac{t \pm x}{\sqrt{2}} \tag{45}$$

Hence, the transformed coordinates of the crossing point $E$ are

$$X_E^+ = \frac{\mu}{\sqrt{2}\kappa}\left(1 - \frac{\mu^2 z}{m_h^2}\right)\ , \qquad X_E^- = \frac{\mu}{\sqrt{2}\kappa}\left(\frac{m_h^2}{\mu^2 z} - 1\right) \tag{46}$$

In this boosted and rotated frame, the equation of the $q_o$ hyperbola of (28) is

$$X^+ = \frac{\mu X^-}{\sqrt{2}\kappa X^- + \mu} \tag{47}$$

And the $\bar{q}_o$ hyperbola of (34) is described by

$$\bar{X}^+ = \frac{-\bar{p}_o^{b+} X^-}{\sqrt{2}\kappa X^- - \bar{p}_o^{b-}} \tag{48}$$

where the bar notation on $\bar{X}^+$ is to distinguish the $\bar{q}_o$ equation from that of $q_o$.

The lower hatched region in Fig. 5 represents $A_{xt}$. To evaluate this area, one may integrate between the hyperbolae curves from $O$ to $E$. Hence,

$$A_{xt} = \int_O^E dX^- \int_{\bar{X}^+(X^-)}^{X^+(X^-)} dX^+ = \int_O^E dX^- \left[X^+\right]_{\bar{X}^+(X^-)}^{X^+(X^-)} = \int_O^E dX^- \left(\frac{\mu X^-}{\sqrt{2}\kappa X^- + \mu} - \frac{-\bar{p}_o^{b+} X^-}{\sqrt{2}\kappa X^- - \bar{p}_o^{b-}}\right) \tag{49}$$

Finishing the last step of integration and using $X_E^-$ from (46) yield the following exact expression for the invariant $A_{xt}$

$$A_{xt} = \frac{1}{2\kappa^2}\left[\frac{m_h^2}{z} - \mu^2 - \mu^2 \log_e\left(\frac{m_h^2}{\mu^2 z}\right)\right] + A.T. \qquad (50)$$

where the first part (with the square brackets) agrees with the Bowler results [12], and the remaining "Additional Term" denoted by the *A.T.*, has the following expression:

$$A.T. = \frac{1}{2\kappa^2}\left[\bar{p}_o^{b+}\mu\left(\frac{m_h^2}{\mu^2 z} - 1\right) + \mu^2 \log_e\left(1 - \frac{\bar{p}_o^{b+}}{\mu}\left(\frac{m_h^2}{\mu^2 z} - 1\right)\right)\right], \qquad (51)$$

making the $A_{xt}$ result exact. Note that the *log* term in *A.T.* is negative and relatively dominant and therefore *A.T.* is always negative (see Fig. 5 for a pictorial representation). The value of the *A.T.* can lead to large or small corrections depending on the c.m. energy. For example, at the c.m. energies near the threshold for the production of a 5.7 *GeV* B-meson containing a 4.7 *GeV b*-quark, *A.T.* contributes 13% to the area, leading to a correction of ~80% to the UCFF fragmentation function of (4). The *A.T.*'s contribution drops at higher c.m. energies due to decreasing values of $\bar{p}_o^{b+}$ given by (31) or (32).

Clearly, although an exact expression for the area is calculated here, the UCLA recursive method [11] used to derive the fragmentation function does not guarantee an exact implementation over the whole c.m. energy spectrum. For instance, the 80% correction from the *A.T.* due to the exact area expression near the B-meson threshold is substantial; however, as discussed in section 2.1, the usefulness of this correction is partially offset by the approximation involved in the fragmentation function at this low energy regime.[2]

As a consistency check, it is also evident from (50) and (51) that when $\mu \to 0$ for the light quarks, $A_{xt} \to m_h^2/2\kappa^2 z = a_{xt}$, as it should.

### *3.3. $Z_{eff}$ – a convenient variable*

Since in (24) $a_{xt} \sim m_h^2/z$, one may find it useful to define a new variable "$z_{eff}$" such that $A_{xt} \sim m_h^2/z_{eff}$. This variable helps with the implementation and conceptually assists with certain facts (see [11] for more detail). In terms of $z_{eff}$, (4) may be written as

[2] However, recall from section 2.1 that the $(1 - m_h^2/Sz)^a$ factor in (4) helps to softens the effect of the approximate nature of the UCFF near threshold, thereby enhancing the impact of the 80% correction.

$$f(z)=\frac{NC^2}{(4\pi)^2}\frac{(1-z)^a}{z}\left(1-\frac{m_h{}^2}{Sz}\right)^{\mathrm{a}}e^{-b\frac{m_h{}^2}{z_{eff}}} \tag{52}$$

This is the UCFF expression written in terms of $z_{eff}$, the analog of $z$ for heavy primary quarks (to be used in the exponent only). More precisely, analogous to (19) one may define

$$z_{eff}\equiv\frac{A_h^{xt}}{A_{xt}}=\frac{m_h^2}{2\kappa^2 A_{xt}} \tag{53}$$

where (22) is utilized for the rightmost result. Furthermore,

$$b=\frac{b'}{2\kappa^2} \tag{54}$$

where the units for $b'$ and $b$ are $fm^{-2}$ and $GeV^{-2}$, respectively. Clearly, $z_{eff}$ is Lorentz invariant because it is only a function of areas. Now, (50) and (53) lead to

$$z_{eff}=\frac{z}{1-\frac{\mu^2 z}{m_h^2}-\frac{\mu^2 z}{m_h^2}\log_e\left(\frac{m_h^2}{\mu^2 z}\right)+\frac{2\kappa^2 z}{m_h^2}\times A.T.} \tag{54}$$

For light quarks, at the limit of $\mu\rightarrow 0$, it is trivial to see that $z_{eff}\rightarrow z$, but not for heavy quarks. For instance, for B-mesons $<z_{eff}>\approx 5<z>$, almost independent of $S$. The rates and distributions for B-mesons differ significantly from data if one uses $z$ rather than $z_{eff}$ in the exponent of UCFF or equivalently uses the light quark space-time area $a_{xt}$ expression instead of $A_{xt}$ expression.

## 4. Comparison to data

The exact expression for the space-time area, $A_{xt}$, given by (50) – which also leads to the exact $z_{eff}$ expression of (54) – has been applied and successfully tested in a recent work [15], while the details of the derivations were being prepared for this manuscript.

After tuning the parameters of the model to the data (which leads to the two major UCFF parameters: $a$ = 1.75, $b$ = 1.10 $GeV^{-2}$ [15], the other four minor parameters unchanged [11], and a best fit $b$-quark mass of 4.7 $GeV$ [15]), in Figs. 6(a–c) a summary of the agreements is shown for the comparisons of all the available meson data in $e^+e^-$ annihilations with the UCLA predictions at the c.m. energies of 90, 29, and 10 $GeV$. A few of the data points in the figure have slightly updated values from [16] since the last comparisons [15]. From Figs. 6(a–c), it is quite evident that the predictions follow the data quite well as the data rates drop with the increasing mass of the states, while in cases such as $K^*$ versus K, $D^*$ versus D, etc., the spin counting effect (via the Clebsch-Gordan coefficient) is evident. These agreements are quite remarkable because they are achieved with the same set of the six tuned model parameters [15], and the comparisons involve data that span a factor of ~600 (~17 to 0.03 per event) in rates, a factor of ~40 (0.135 to 5.5 *GeV*) in hadron mass, a factor ~9 (10 to 91 *GeV*) in c.m. energy, all five accessible quark flavors, and several spin states, including orbitally excited charmed ($D^{**}$) and bottom ($B^{**}$) mesons.

Statistical analyses based on the same data comparisons reaffirm that STAL can function as the single dominant physical basis for hadronization − requiring no additional physical process to account for the observed data. More quantitatively, it is shown that the combined effects of all secondary processes present beyond the STAL, if any, contribute less than 20% to the rates [15]. With more accurate future data this limit may improve.

Baryon modeling results are encouraging [11,15]. However some subtleties in their implementation, such as correlations and "popcorn" structures [11] need further investigation.

In calculating the $A_{xt}$ space-time area, if the hyperbolic world-lines of the heavy primary quarks are ignored and correction to the space-time area is not made due to this mass related effect, one finds large deviations from data in the B meson sector; i.e., for the rates, a $\chi^2$/DOF of 7.7 for the latter, compared to 1.3 when the $A_{xt}$ area expression of (50) is used. Furthermore, reference [15] shows significant deviations in the fits to the energy spectra when this quark mass effect is ignored.

## 5. Summary

Space-time area laws are instrumental in providing dominant physical bases for strongly interacting phenomena where perturbation techniques fail. Hadronization phenomenon, being primarily based on non-perturbative QCD, is one such case. For instance, in the UCLA scheme, a space-time area law ("STAL") is shown to work as the underlying basis for hadronization. An exact expression for the relevant space-time areas involved in the final state hadronic events is derived and used in the UCLA hadronization fragmentation function. With this implementation, all recent meson data from $e^+e^-$ collisions at all available c.m. energies appear remarkably consistent with the idea that the STAL assumption is a valid QCD-motivated basis for hadronization. In addition, the derived area expression unifies the treatment of light and heavy quarks in the modeling.

Equipped with the STAL principle and the derived space-time area expression, the UCLA approach may be regarded as a first step toward establishing a pathway from non-perturbative QCD to the hadronization data.

**Figure captions**

**Fig. 1.** (a) An energetic quark-antiquark pair ( $q\overline{q}$ ) is produced at point *O*, for instance, from decay of a $\gamma^*$ or $Z^o$ from an $e^+e^-$ annihilation. Small open and closed circles are quarks and antiquarks, respectively, and the dashed lines connecting them represent the strings (color flux tubes), and the solid lines are light quark world-lines which are on the light-cone. As the time evolves vertically upward on the page, the string between $q$ and $\overline{q}$ stretches which subsequently breaks at the first vertex point, producing a new quark-antiquark pair, $q'\,\overline{q}'$. Later, at other vertices, other new pairs are produced. (b) From breakings of the color fields, several vertices may be produced leading to pairs joined by new strings, such as $q_o\overline{q}_1$ and $q_1\overline{q}_o$, which subsequently can break into other pairs. Finally a stage is reached (line segments in the upper part of the figure) where similar pairs are confined inside hadrons via stable yo-yo mode motions (shown by shaded rectangles). Baryons are not shown, however if any break point contains a diquark pair, it can lead to baryon production. The total event space-time area marked by $A_{\mathrm{plane}}$ (bordered by the solid line segments) and $a_{\mathrm{xt}}$ (hatched area), are related to the event and hadron formation probabilities, respectively.

**Fig. 2.** While massless quarks travel on the light-cone (dotted lines), heavy primary quark's classical world-lines in 1+1 dimensions are curved (solid curves) and they modify the space-time areas. $A_{\mathrm{xt}}$ (hatched region) and $A_{\mathrm{plane}}$ (total event area, bordered by the solid line segments and curves) are modified by the hyperbolic world-lines of heavy primary quarks. Therefore, event and hadron formation probabilities are also modified.

**Fig. 3.** A stable meson: as a square in its c.m. frame and as a rectangle in a boosted frame, while the constituent quark and antiquark are confined in a yo-yo mode motion. The shape of the figure is the same (with the exception of units and scale) in *xt*, momentum, light-cone *xt*, and light-cone momentum frames. For instance, if the coordinates of the turning point *Q* is $(x_q, t_q)$ in *xt* frame, it is $(x_q^+ = x_q + t_q, x_q^- = t_q - x_q)$ in $x^+x^-$ frame, $(p_{oq} = \kappa t_q, E_{oq} = \kappa x_q)$ in $pE$ system, and $(p_{oq}^+ = \kappa x_q^+, p_{oq}^- = \kappa x_q^-)$ in $p^+p^-$ frame (section 3.1). Note that, the momentum coordinates are meant to represent initial values of parameters, while *xt* coordinates show current parameter values.

Both the square and the rectangle have the same areas, and in *xt* space the area is related to that in momentum space by $S_{xt} = S_p/\kappa^2$, with $\kappa \approx 1$ *GeV/fm*.

**Fig. 4.** The right side of Fig. 1(b), shown in light-cone momentum space.

**Fig. 5.** The hyperbolic world-lines of the heavy quarks $q_o$ and $\bar{q}_o$ in a boosted frame where $q_o$ was initially at rest. The lower hatched area $A_{xt}$ (a Lorentz invariant) is shown, which is bounded by the $q_o$ and $\bar{q}_o$ curved paths and the light-cone path segment of $\bar{q}_1$ and its extension.

**Fig. 6.** At c.m. energies of (a) 91 *GeV*, (b) 29 *GeV*, and (c) 10 *GeV*, all available data rates for mesons produced in $e^+e^-$ collisions [16] are compared with the UCLA model predictions in terms of the number of mesons produced per event. Particle names and mass value labels for each entry are displayed in these plots. Model predictions are shown even if usable data do not exist.

See Figures in the next few pages:

**Fig. 1**

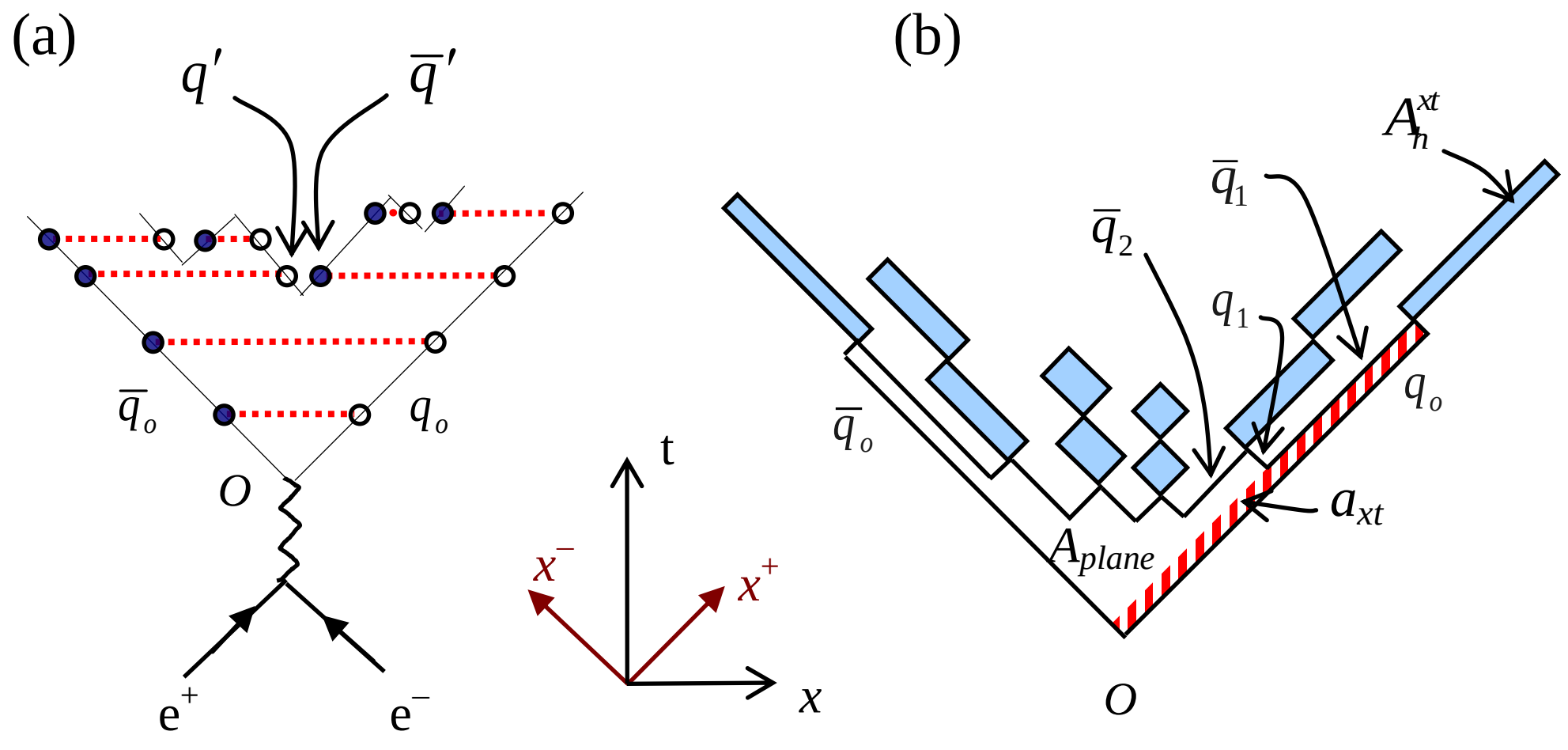

(a)
(b)
$q'$
$\bar{q}'$
$\bar{q}_o$
$q_o$
$O$
e$^+$
e$^-$
t
$x^-$
$x^+$
$x$
$A_h^{xt}$
$\bar{q}_1$
$\bar{q}_2$
$q_1$
$\bar{q}_o$
$q_o$
$a_{xt}$
$A_{plane}$
$O$

**Fig. 2**

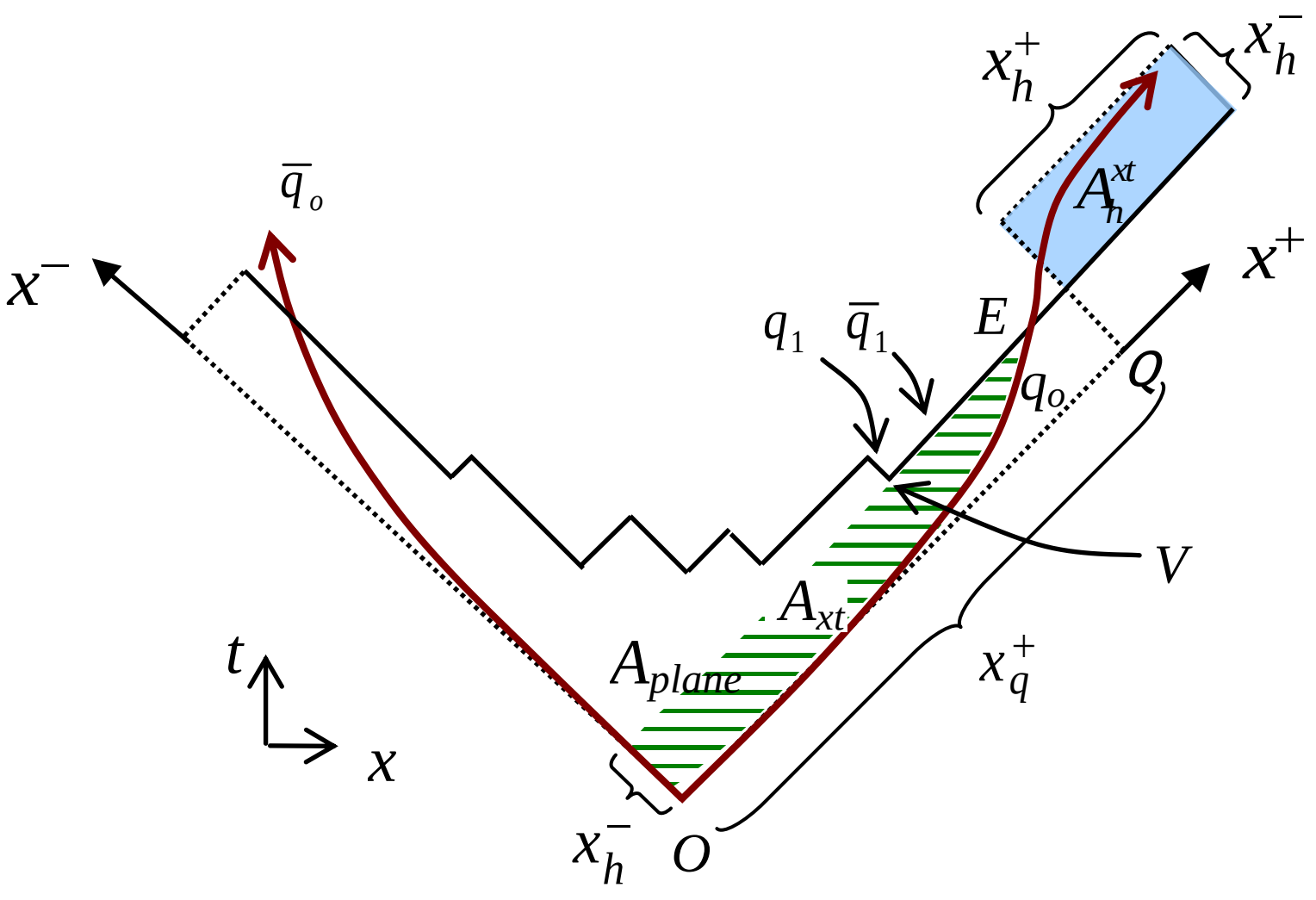

**Fig. 3**

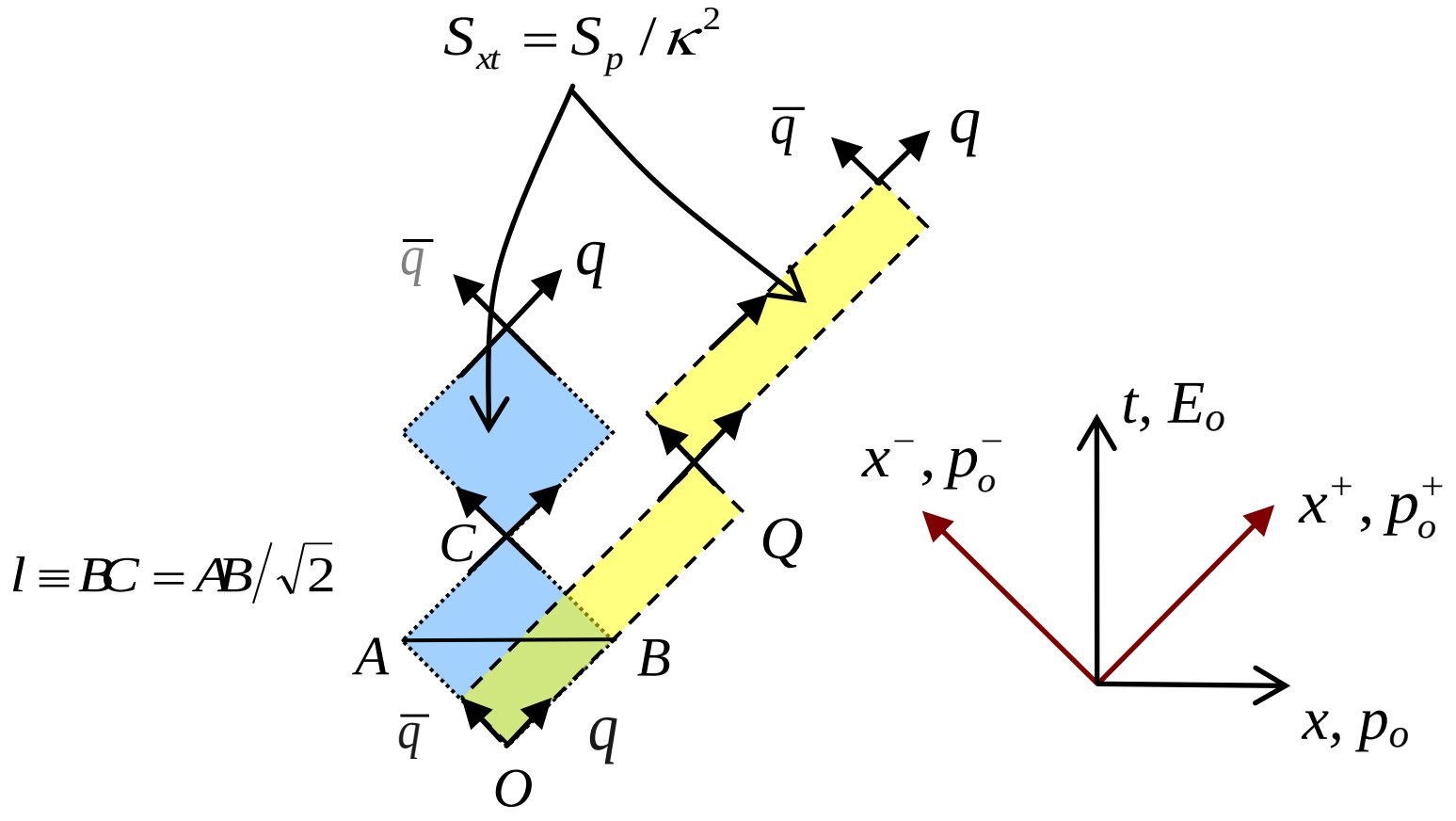

$S_{xt} = S_p / \kappa^2$
$\bar{q}$
$q$
$\bar{q}$
$q$
$t, E_o$
$x^-, p_o^-$
$x^+, p_o^+$
$C$
$Q$
$l \equiv BC = AB/\sqrt{2}$
$A$
$B$
$x, p_o$
$\bar{q}$
$q$
$O$

**Fig. 4**

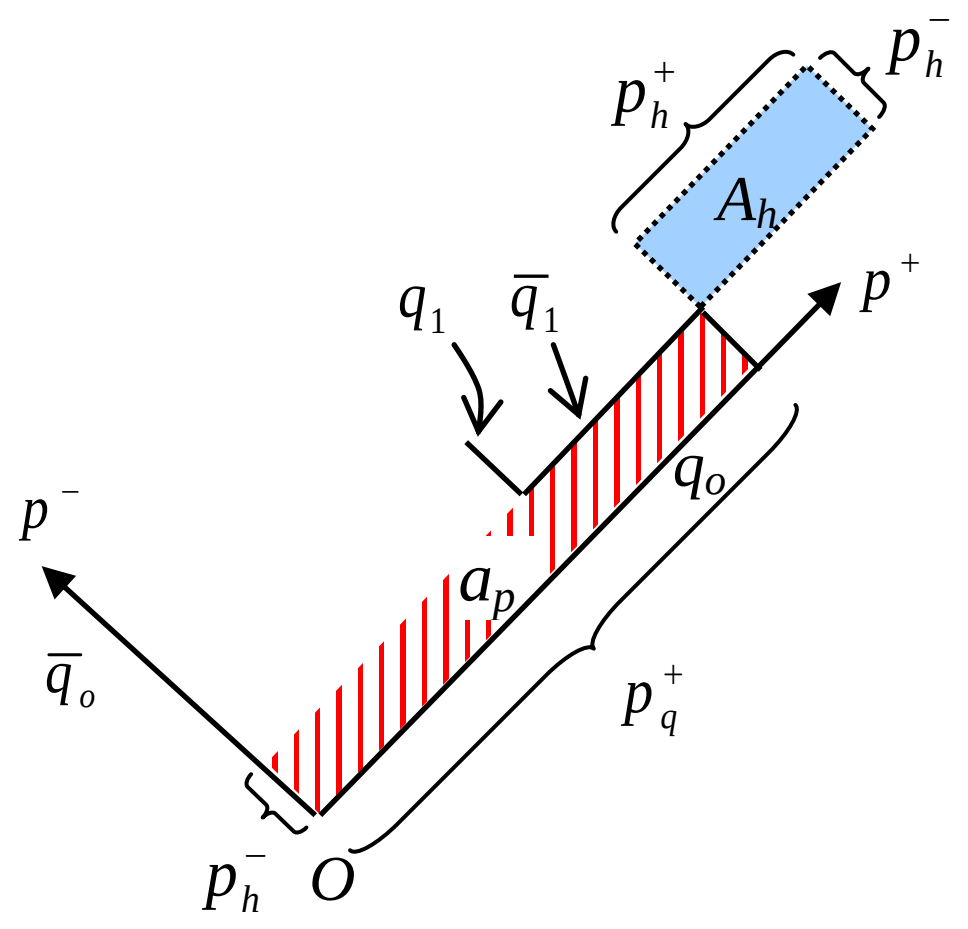

**Fig. 5**

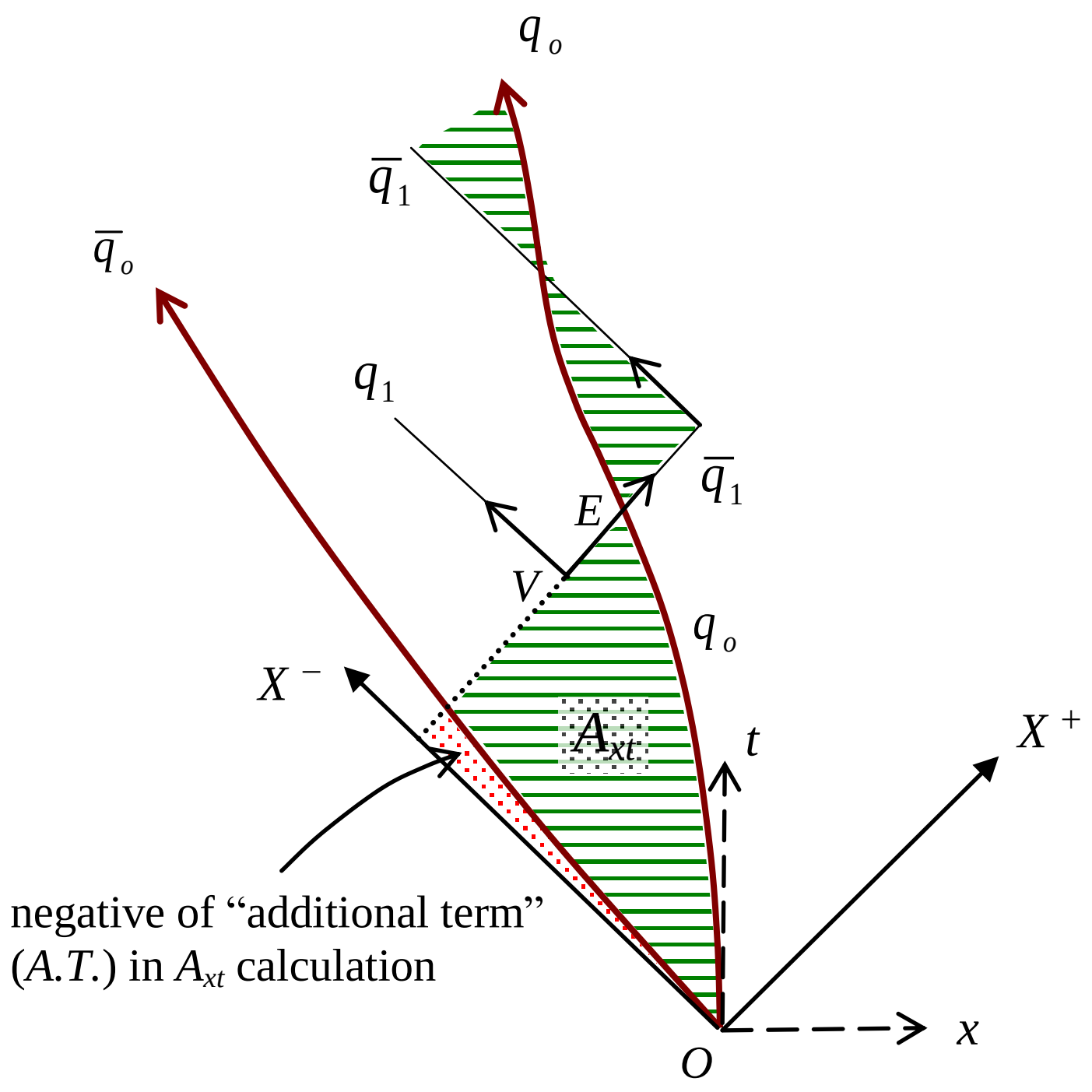

$q_o$
$\bar{q}_1$
$\bar{q}_o$
$q_1$
$E$
$\bar{q}_1$
$V$
$q_o$
$X^-$
$A_{xt}$
$t$
$X^+$
negative of "additional term"
($A.T.$) in $A_{xt}$ calculation
$x$
$O$

**Fig. 6**

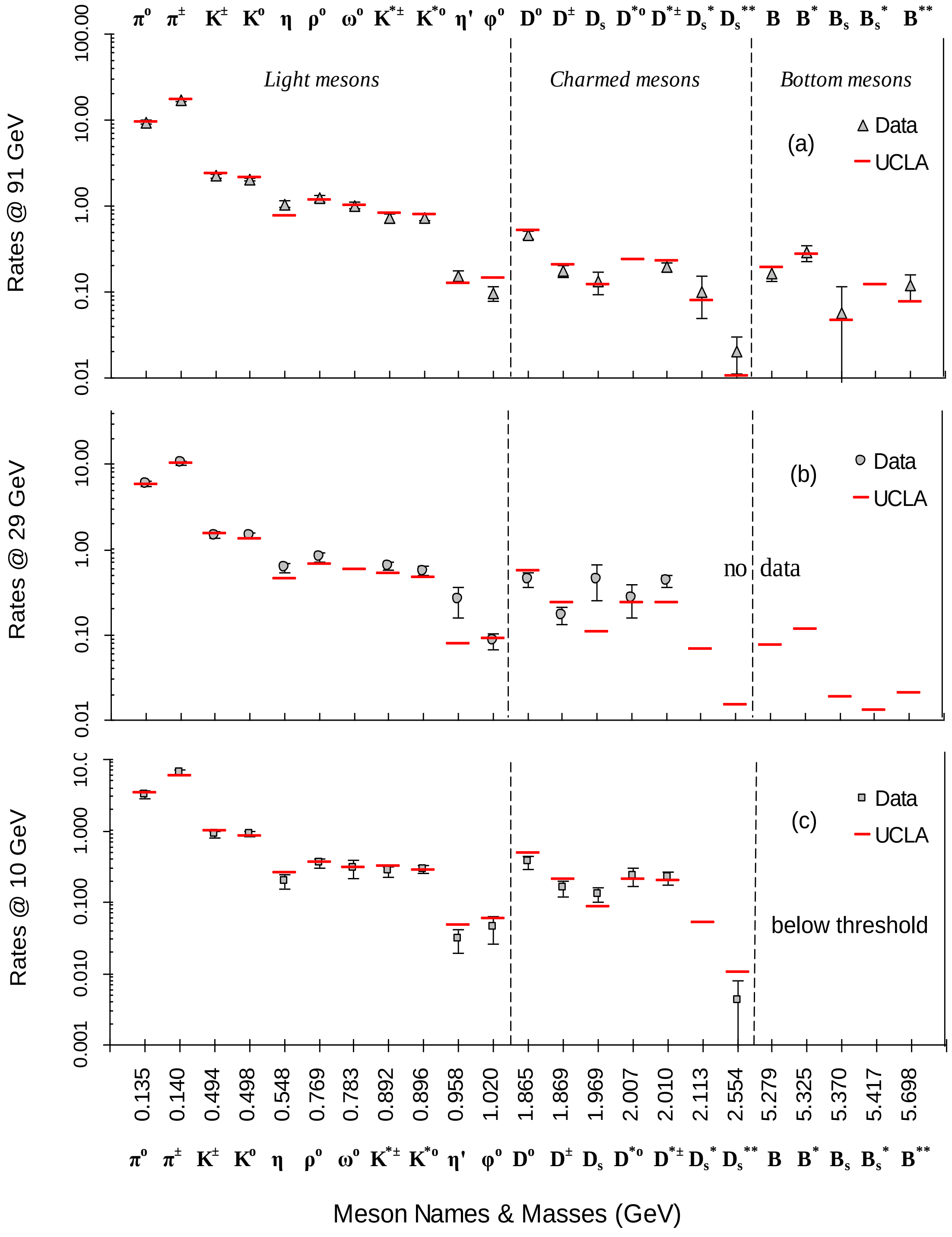
Light mesons
Charmed mesons
Bottom mesons
Data
UCLA
(a)
(b)
(c)
no data
below threshold
Rates @ 91 GeV
Rates @ 29 GeV
Rates @ 10 GeV
Meson Names & Masses (GeV)